\documentclass[twocolumn,notitlepage,nofootinbib,amsmath,amssymb,aps,prl]{revtex4-1}

\usepackage[T1]{fontenc}
\usepackage{graphicx}
\usepackage{dcolumn}
\usepackage{bm}

\usepackage{color}

\begin{document}

\title{On parametric tests of relativity with false degrees of freedom}

\author{Alvin J.\ K.\ Chua}
\email{Alvin.J.Chua@jpl.nasa.gov}
\affiliation{Jet Propulsion Laboratory, California Institute of Technology, Pasadena, CA 91109, U.S.A.}

\author{Michele Vallisneri}
\email{Michele.Vallisneri@jpl.nasa.gov}
\affiliation{Jet Propulsion Laboratory, California Institute of Technology, Pasadena, CA 91109, U.S.A.}

\date{\today}

\begin{abstract}
General relativity can be tested by comparing the binary-inspiral signals found in LIGO--Virgo data against waveform models that are augmented with artificial degrees of freedom. This approach suffers from a number of logical and practical pitfalls. 1) It is difficult to ascribe meaning to the stringency of the resultant constraints. 2) It is doubtful that the Bayesian model comparison of relativity against these artificial models can offer actual validation for the former. 3) It is unknown to what extent these tests might detect alternative theories of gravity for which there are no computed waveforms; conversely, when waveforms are available, tests that employ them will be superior.
\end{abstract}

\maketitle


\paragraph{Introduction.}

The detection of unexpectedly loud gravitational waves (GWs) from inspiraling binary black holes \cite{PhysRevX.9.031040} has allowed precise tests of the general-relativistic (GR) predictions for the shape of these signals \cite{PhysRevD.100.104036}. In a \emph{parametric} test, one expands the GR model by introducing a functional dependence on one or more parameters that describe additional degrees of freedom (DOFs), then measures them along with the astrophysical source parameters. The base GR model is recovered for specific values of the new parameters; estimates that are significantly displaced from these values would signal a violation of Einstein's theory. We will call the additional DOFs \emph{true} if they describe quantities with independent physical meaning (such as the hypothetical mass of the graviton), or \emph{false} if they extend possible signal shapes in directions that are not motivated by physics. There is a gray area between the two, inhabited for example by models that modify individual coefficients in the frequency-domain post-Newtonian (PN) expansion of the GW phase \cite{Arun_2006,PhysRevD.74.024006,PhysRevD.80.122003}, since the dominant contribution from an alternative theory of gravity can often be mapped to a particular PN coefficient (\cite{PhysRevD.94.084002} and references within).

Many studies in the GW literature involve DOFs that are false, or not unequivocally true (e.g., \cite{PhysRevD.82.064010,PhysRevD.84.062003,PhysRevD.87.102001,PhysRevD.89.064037,PhysRevD.85.082003,Li_2012,PhysRevD.89.082001,PhysRevD.97.044033,PhysRevD.99.124044,PhysRevLett.123.121101}). These studies are occasionally affected by a number of statistical misconceptions that invalidate or weaken some of their claims. First, tests with different parameters are sometimes compared in terms of their ``stringency'', i.e., the fractional precision of estimation for the testing parameters: doing so is largely meaningless, because the precision is not parametrization-invariant. Second, Bayes factors for GR against false-DOF theories are claimed to validate the former: this is problematic both logically, due to the dubious standing of the artificial theories, and practically, because Bayes factors are influenced by parametrization and the arbitrary specification of priors. Third, false-DOF tests are seen as generally diagnostic of unspecified modified-gravity theories, but this generality may only be established if one can already produce waveforms from those theories---in which case one is better off directly using those waveforms to begin with.

We will examine each of these arguments in turn, using a toy analogy inspired by a recent false-DOF test on the GW sources detected by LIGO--Virgo \cite{haster2020pi}. In this innovative test, the mathematical constant $\pi$, which occurs formally in several of the PN phase coefficients, is elevated to a source-independent parameter that is estimated alongside the physical parameters. (More banal appearances of $\pi$ arising from the stationary phase approximation are left undisturbed.) As discussed above, a violation of GR would be identified by recovering a value of $\pi$ that is statistically distinct from its generally accepted baseline,\footnote{For different baselines, see \cite{singmaster1985legal}.} after measurement errors are taken into consideration. Ref.~\cite{haster2020pi} includes claims of the variety listed above: that the variable-$\pi$ test provides the most stringent constraint to date on the PN coefficients; that it validates GR by way of a large Bayes factor between the base theory and its variable-$\pi$ extension; and that it can robustly indicate the presence of GW phasing effects due to generic non-GR theories.

\paragraph{On the stringency of constraints.}

As an analogy, consider the hypothesis that a set of $N$ round two-dimensional shapes are perfect circles, in accordance with some General Theory of Circularity. Three noisy measurements are made on each shape: the radius $r$, the circumference, and the area. The forward model for the shape observables in General Circularity is thus
\begin{equation}
\mathrm{circ}(r)=\left(r,2\pi r,\pi r^2\right).
\end{equation}
For simplicity, assume that both the relative strength of noise in the three observables and the total strength of noise are known for each shape, such that GW-analogous versions of noise whitening and marginalization over signal-to-noise ratio are trivial to perform. (Our conclusions would not be changed by treating noise in full generality.) Now suppose the shapes are indeed circles; the measured data over $N$ shapes is taken to be
\begin{equation}
d_\mathrm{circ}=\{\rho_i\,\mathrm{circ}(r_i)+n_i\,|\,i=1,\ldots,N\},
\end{equation}
with $\rho_i\sim\mathcal{U}(1,10)$, $r_i\sim\mathcal{U}(1,3)$, and $n_i \sim \mathcal{N}(0,I_3)$.

\begin{figure}[!tbp]
\centering
\includegraphics[width=\columnwidth]{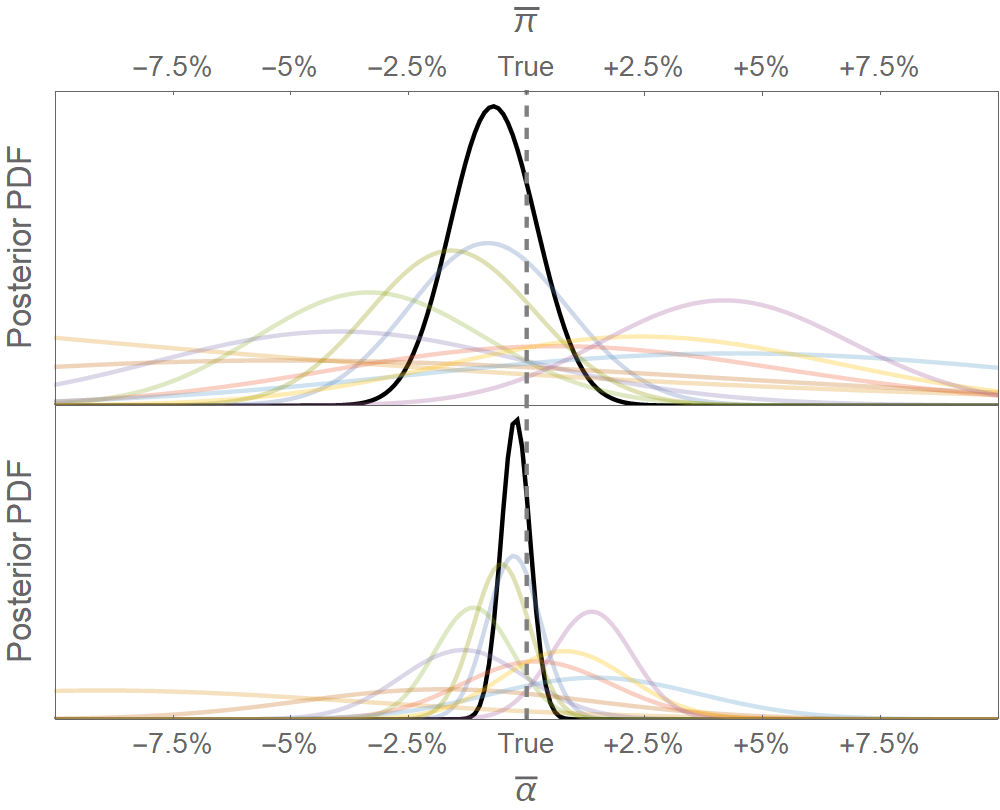}
\caption{Individual (colored) and joint (black) posteriors for $\bar{\pi}$ and $\bar{\alpha}$, as estimated from a sample $d_{\mathrm{circ}}$ data set with $N=10$.}
\label{fig:stringency}
\end{figure}

In the spirit of \cite{haster2020pi}, define the variable-$\pi$ model as
\begin{equation}
\mathrm{circ}_{\bar{\pi}}(r,\bar{\pi})=\left(r,2\bar{\pi} r,\bar{\pi} r^2\right),
\end{equation}
where $\bar{\pi}$ denotes the false DOF, and $\mathrm{circ}_{\bar{\pi}}(r,\pi)=\mathrm{circ}(r)$. Given the data set $d_\mathrm{circ}$, a likelihood function of $(r,\bar{\pi})$ may be constructed in the usual way with $\mathrm{circ}_{\bar{\pi}}$ and the Gaussian noise assumption. Take the true distribution of $r_i$ as the prior for $r$, and $\bar{\pi}\sim\mathcal{U}(\pi-10\%,\pi+10\%)$ as an arbitrary prior for $\bar{\pi}$. The joint posterior density $p(\bar{\pi}|d_\mathrm{circ})$ is then computed by marginalizing over $r$ separately for each shape, while treating $\bar{\pi}$ as a ``universal'' parameter that is common to all shapes. The top panel of Fig.~\ref{fig:stringency} shows $p(\bar{\pi}|d_\mathrm{circ})$ for a sample data set with $N=10$, together with the posteriors from the individual shapes. A joint measurement of $\bar{\pi}=3.119^{+0.048}_{-0.047}$ is obtained from $p(\bar{\pi}|d_\mathrm{circ})$ (maximum a posteriori, with errors corresponding to the $5\%$ and $95\%$ quantiles); it is consistent with $\pi$, and thus appears to validate General Circularity.

But how should we regard the stringency of the $\bar{\pi}$ constraint with respect to other tests of the theory? Consider an alternative variable-``one'' model
\begin{equation}
\mathrm{circ}_{\bar{\alpha}}(r,\bar{\alpha})=\left(r,2\pi\bar{\alpha}^3r,\pi\bar{\alpha}^3r^2\right),
\end{equation}
where $\bar{\alpha}$ denotes ``one'',\footnote{As it does in many ancient Greek manuscripts \cite{ifrah2000universal}.} and $\mathrm{circ}_{\bar{\alpha}}(r,1)=\mathrm{circ}(r)$. Repeating the test of General Circularity as above with the prior $\bar{\alpha} \sim \mathcal{U}(1-10\%,1+10\%)$ yields a fractional constraint that is thrice as stringent\footnote{The same fractional precision can be obtained by estimating $\sqrt[3]{\bar{\pi}}$ instead of $\bar{\pi}$, at the cost of pithiness.} (see Fig.~\ref{fig:stringency}). Indeed, one may manufacture tests of arbitrary precision simply by increasing the power of $\bar{\alpha}$. In the same way, inserting $\bar{\alpha}$ with some exponent $>1$ wherever $\pi$ occurs formally in the PN coefficients will be enough to give a more stringent false-DOF constraint than that reported in \cite{haster2020pi}.

Takeaway point: The fractional precision of false-DOF determination is not parametrization-invariant, so it cannot be used to assess the ``stringency'' of a test of GR. By contrast, precision \emph{is} relevant for the estimation of true DOFs, for which there may be predicted values or constraints from other theories or experiments (e.g., the graviton mass affects both GW propagation and the gravitational potential in Solar-System measurements).

\paragraph{On the abuse of Bayes factors.}

Bayesian model selection is a common feature in parametric tests of GR. In the archetypal false-DOF test, the Bayes factor may be used to compare GR against an expanded model in which a finite set of PN coefficients are allowed to vary fractionally. One notable variant \cite{PhysRevD.85.082003,Li_2012,PhysRevD.89.082001} instead considers the set of submodels defined by the power set of these coefficients, and pits GR against the logical sum of all submodels (this test does not appear to have been performed on actual data yet). In \cite{haster2020pi}, the Bayes factor between GR and its variable-$\pi$ extension is computed using the joint data from all compact binary coalescences reported so far; with the prior $\bar{\pi}\sim\mathcal{U}(-20,20)$ (our notation), a Bayes factor of $\approx 300$ is found in favor of GR. Unfortunately, such calculations all bear little import to our beliefs about the correct theory of gravitation, for two main reasons. These are outlined below for the variable-$\pi$ test, but apply equally to any false-DOF test of GR.

First: it has not been demonstrated that a variation of $\pi$ in the PN phasing maps, even approximately, to any theoretically viable modification of GR. (This goes for the fractional variation of PN coefficients as well.) Thus a fully specified theory is compared with a mathematical artifice that, by all rights, should have vanishing prior belief. Recall that Bayesian model selection relies on the posterior odds, which equal the Bayes factor times the prior odds; one then wonders if \emph{any} Bayes factor would convince us that GR is violated, instead of merely casting suspicion on the data, the priors, etc. It might be argued that the identification of such systematics is indeed a purpose of the test. In that case, the test should be interpreted as validating not so much GR, but rather data collection and analysis under the GR hypothesis.

Second: if GR is correct, the variable-$\pi$ model can fit the data only marginally better (by overfitting noise), but is penalized by a large Occam factor because $\bar{\pi}$ needs to be fine-tuned towards the ``true'' value within its prior bounds. For a false DOF, these bounds are inevitably quite arbitrary. In the case of our toy analogy and example data, the Bayes factor in favor of General Circularity is $6.5$ with variable $\pi$ and $19.3$ with variable ``one'', indicating that the factor itself is not parametrization-invariant. Furthermore, if the prior interval for each variable is expanded from $\pm 10\%$ to $\pm 25\%$, the Bayes factor simply increases by the same proportion since the relative width of the joint posterior becomes $2.5$ times tighter around the true value (and since no significant secondary maxima exist, as is likely in the GW case with $N$ large).

Takeaway point: Claims that GR is validated by large Bayes factors against false-DOF extensions of GR should be taken lightly---the factors in such tests have a dubious interpretation, and they can also be inflated or deflated by arbitrary analysis choices.

\paragraph{On the intersection of models.}

Waveform models are not available for many theories of modified gravity, and little is known except that they are likely to manifest as changes in the PN coefficients. False-DOF tests then extend the manifold of possible signals along directions (in data space) that may or may not be motivated by physics, in the hope that it will intersect significantly with putative models in actual modified theories. If the additional DOFs are not orthogonal to the physical DOFs, the two sets become entangled within Bayesian inference, creating the possibility of fundamental bias (i.e., modified gravity biases the recovery of physical parameters \cite{PhysRevD.80.122003}) and stealth bias (biased physical parameters hide modified gravity \cite{PhysRevD.87.102002}). Note, however, that it is possible to determine the extent of intersection between a false-DOF theory and a specific modified theory \emph{only} if waveforms from the latter are on hand---but then one may as well use those to perform a more informative test. Absent that determination, the statistical interpretation of the false-DOF test remains ambiguous.


In terms of our toy analogy, assume now that the observed shapes are in truth ellipses---such that General Circularity no longer describes the data accurately, and the true theory of roundness is Ellipticity. Using Ramanujan's second approximation \cite{ramanujan1914modular} to the circumference of an ellipse, define the modified-roundness model
\begin{equation}
\mathrm{ell}(a,e)=\left(a,\pi(a+b)\left(1+\tfrac{3h}{10+\sqrt{4-3h}}\right),\pi ab\right),
\end{equation}
where $b=a\sqrt{1-e^2}$, $h=(a-b)^2/(a+b)^2$, and $\mathrm{ell}(r,0)=\mathrm{circ}(r)$. The measured data over $N$ shapes is taken to be
\begin{equation}
d_{\mathrm{ell}}(e)=\{\rho_i\,\mathrm{ell}(r_i,e)+n_i\,|\,i=1,\ldots,N\},
\end{equation}
with the same realization of $\{\rho_i,r_i,n_i\}$ as used in Fig.~\ref{fig:stringency}. For $e \gtrsim 0.75$, the variable-$\pi$ test is able to ``detect'' the presence of eccentricity; specifically, the joint posterior density $p(\bar{\pi} | d_\mathrm{ell})$ is peaked away from $\bar{\pi} = \pi$, and the Bayes factor for the $\mathrm{circ}$ model over the $\mathrm{circ}_{\bar{\pi}}$ model is $<1$. Likewise in \cite{haster2020pi}, an experiment with simulated data suggests that the variable-$\pi$ test can indicate a finite value for the Compton wavelength of the graviton.

Focus, however, on the statistical significance of observing an anomalous posterior density $p(\bar{\pi}|d)$. An occurrence of the null (true) value deep into the tails of the posterior does not map directly to the detection of an anomaly with high statistical significance. This is because the \emph{coverage} of Bayesian credible intervals is only guaranteed for a population that is distributed according to the prior used in the Bayesian analysis. For example, one might expect that $p(\bar{\pi} < \pi|d) < 0.01$ would occur less than $1\%$ ``of the time'', i.e., in $<1\%$ of similar experiments. The only way this statement would be a certainty is if those data sets are generated using the $\mathrm{circ}_{\bar{\pi}}$ model with the analysis priors for $(r,\bar{\pi})$, or the $\mathrm{ell}$ model with priors for $(a,e)$ that map to the analysis priors (the latter is not even assured to be always possible).

The reliability of a test of GR may nevertheless be validated by a frequentist analysis of its performance as a detection scheme \cite{PhysRevD.86.082001}. One determines the probability of a false GR-violation claim if GR is true (a.k.a. a Type-I error), and of a false dismissal if an alternative theory of gravity is correct (a Type-II error). Type-I errors seem more pertinent here, since it would be far more egregious to claim erroneously that GR has been falsified, than to miss out on detecting modified gravity; furthermore, Type-II errors are quantified only for individual modified theories, and only when their waveform models are available. Still, it is clear that a parametric test will not be particularly useful if it often incurs Type-II errors on most of the modified theories that can be tested.

\begin{figure}[!tbp]
\centering
\includegraphics[width=\columnwidth]{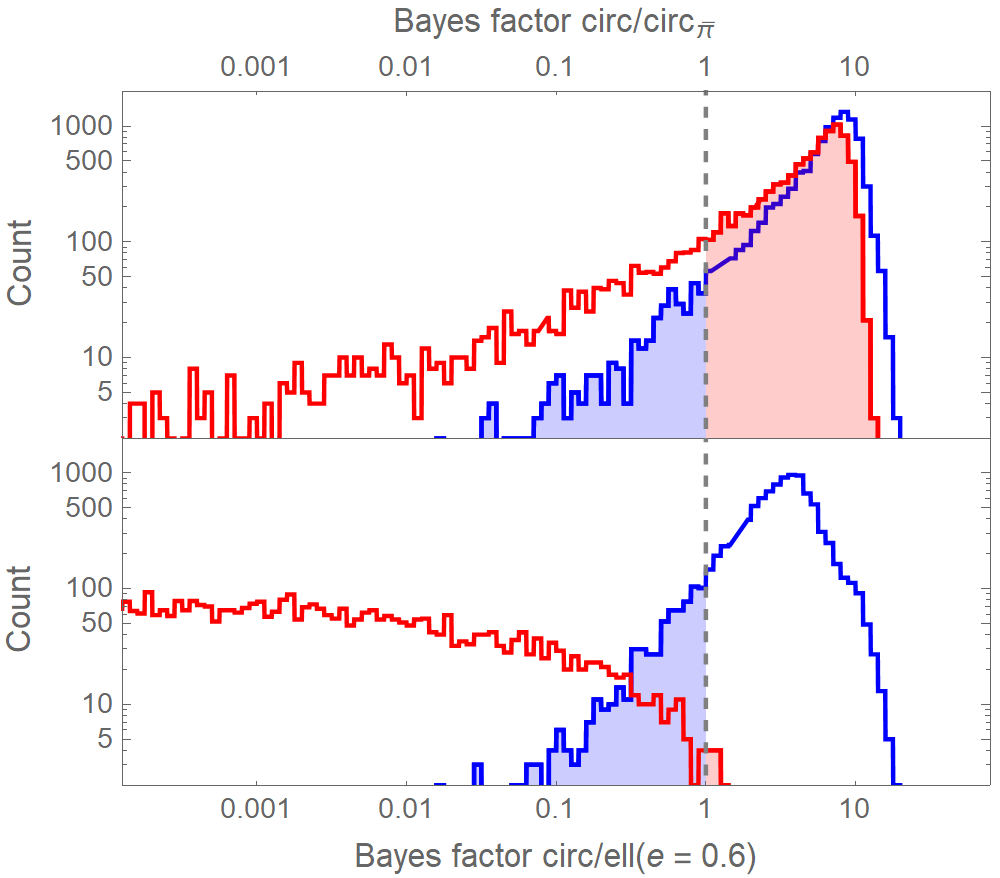}
\caption{Distributions of $\mathrm{circ}$-over-$\mathrm{circ}_{\bar{\pi}}$ and $\mathrm{circ}$-over-$\mathrm{ell}$ Bayes factors from $10^4$ realizations of $d_{\mathrm{circ}}$ (blue) and $d_{\mathrm{ell}}(0.6)$ (red). Shaded blue areas on the left of unity correspond to Type-I errors; shaded red areas on the right to Type-II errors. Logarithmic vertical axes are used to emphasize the tails.}
\label{fig:bferrors}
\end{figure}

As it turns out, the variable-$\pi$ test has limited discriminative power when testing General Circularity against the modified-roundness theory of Ellipticity, using the Bayes factor as a detection statistic with the natural threshold\footnote{The threshold may of course be adjusted to attain a desired false-alarm rate, without changing the conclusions of the exercise.} of unity. Consider the case $e=0.6$, which the test is unable to detect for the given realization of $\{\rho_i,r_i,n_i\}$, with a reported $\mathrm{circ}$-over-$\mathrm{circ}_{\bar{\pi}}$ Bayes factor of $7.7$. This Type-II error is not a rare event; from a simulated population of $10^4$ data sets $d_{\mathrm{ell}}(0.6)$ (each with $N=10$ but different realizations of $\{\rho_i,r_i,n_i\}$), the probability of obtaining a Bayes factor $>1$ is $84.2\%$. Thus the variable-$\pi$ test is sensitive to an eccentricity of $0.6$ in just $15.8\%$ of experiments. It does fare better on Type-I errors, with a false-alarm rate of $3.6\%$ on $10^4$ null data sets $d_{\mathrm{circ}}$ (using the same $\{\rho_i,r_i,n_i\}$ as above). The distributions of the detection statistic are shown in the top panel of Fig.~\ref{fig:bferrors}. For comparison, one may repeat this analysis with the true model $\mathrm{ell}$ in place of $\mathrm{circ}_{\bar{\pi}}$, and using $e$ as the testing variable with a full prior $e\sim\mathcal{U}(0,1)$. This yields a comparable Type-I error rate ($6.8\%$), but a hugely improved sensitivity ($99.9\%$; or $99.7\%$ if the false-alarm rate is matched to that of the variable-$\pi$ test).

Takeaway point: The utility of a false-DOF test of GR can only be assessed with respect to specific modified-gravity models---by evaluating the degeneracy between the physical and testing parameters, and by characterizing Type-I/II errors in a detection scheme. Without such validation, the statistical significance of test outcomes is unspecified, so there might be no good reason for the test to be performed at all.

\paragraph{Conclusion.}

False-DOF tests compare GR to strongly specified, unphysical theories that select arbitrary directions within (and likely \emph{beyond}) the signal space spanned by viable theories of modified gravity. That the tests might favor GR is as much a statement on the invalidity of the artificial theories as it is on the validity of relativity. Caution is needed in the interpretation of any quantitative results, since the tests are not invariant with respect to parametrization and prior specification. A principled statistical assessment of each test is only possible against individual modified-gravity theories for which waveforms are available, in which case theory-specific tests will almost certainly be superior.

\paragraph{Acknowledgements.}

We thank Maria Charisi for conferring on Greek numerals. This work was supported by the Jet Propulsion Laboratory (JPL) Research and Technology Development program, and was carried out at JPL, California Institute of Technology, under a contract with the National Aeronautics and Space Administration. \copyright\,2020 California Institute of Technology. U.S.\ Government sponsorship acknowledged.

\bibliography{main}

\end{document}